\documentclass[12pt]{article}
\usepackage[dvips]{graphicx}
\usepackage{epsfig}
\usepackage{amsmath}
\usepackage{bm}
\usepackage{times}
\usepackage{hyperref}
\topmargin=\paperheight
\advance\topmargin by -\headheight
\advance\topmargin by -\headsep
\advance\topmargin by -\textheight
\advance\topmargin by -\footskip
\divide\topmargin by 2
\advance\topmargin by -1 in
\makeatletter
\@addtoreset{equation}{section}
\renewcommand\theequation{\thesection.\@arabic\c@equation}
\makeatother

\newcommand{\calo}{{\cal O}}

\def\roughly#1{\mathrel{\raise.3ex\hbox{$#1$\kern-.75em%
\lower1ex\hbox{$\sim$}}}}

\begin{document}
\begin{titlepage}
\begin{center}
{\LARGE\bf  Chiral Extrapolation of the Strangeness Changing  \\[2mm] 
 Scalar {\boldmath$K \pi$} Form Factor}
\\[12mm]
{\normalsize\bf V\'eronique Bernard~${}^{a,}$\footnote {Email:~bernard@ipno.in2p3.fr}
and
Emilie Passemar~${}^{b,}$\footnote{Email:~emilie.passemar@ific.uv.es} 
}\\[4mm]

{\small\sl ${}^{a}$Groupe de Physique Th\'{e}orique, IPN,
} \\
{\small\sl Universit\'{e} de Paris Sud-XI/CNRS , F-91406 Orsay, France}\\

{\small\sl ${}^{b}$ Departament de F{\'i}sica Te{\`o}rica, IFIC, Universitat de Val\`encia - CSIC,\\
Apartat de Correus 22085, E-46071 Val\`encia, Spain}\\
[12mm]
\end{center}

\vspace{2cm}
\noindent{\bf Abstract:}
\noindent
We perform a chiral extrapolation of lattice data on the scalar $K \pi$ form factor and the
ratio of the kaon and pion decay constants within Chiral Perturbation Theory to two loops.
We determine the value of the scalar form factor at zero momentum transfer, at the Callan-Treiman
point and at its soft kaon analog as well as its slope. Results are in good agreement with
their determination from experiment using the standard couplings of quarks to the W boson. The
slope is however rather large. A study of the convergence of the chiral expansion is also performed.
\end{titlepage}
\pagenumbering{arabic}
\renewcommand{\thefootnote}{\arabic{footnote}}
\parskip12pt plus 1pt minus 1pt
\topsep0pt plus 1pt
\section{Introduction}
In recent years lots of progress has been made in QCD lattice calculations.
One important progress in the light quark sector concerns the values of
the quark masses that can now be reached. These are very close to the physical
ones  making  a controlled, i.e. trustable, chiral extrapolation of the lattice results
to the physical points possible. A very powerful model-independent framework 
to perform this extrapolation is Chiral Perturbation Theory (ChPT), the 
Effective Field Theory of QCD at low energies. Indeed it allows to calculate
low-energy QCD processes in terms of the light pseudoscalar mesons masses.  
Hadron properties are presently actively  studied on
the lattice and chiral extrapolations  to their physical values 
are performed, see for example \cite{sh06,mesc07,be07}.

At the same time lots of effort is put into testing the 
Standard Model (SM). In order to do so one  has to have very precise 
determinations 
of the QCD quantities which generally enter the different processes under consideration.  
Two very interesting quantities in this respect are the strangeness changing scalar $f_0$ 
and vector $f_+$ form factors which are measured in $K_{l3}$ decays \cite{Antonelli:2008jg}. 
Indeed a measurement of the $K_{\ell3}$ inclusive decay rate leads 
to the extraction of the product of the vector form factor at zero momentum transfer $f_+(0)$  
and of the CKM matrix element $|V_{us}|$. Consequently the knowledge of  $f_+(0)$ allows
to extract this matrix element and thus to test the unitarity relation
between the elements of the first row of this matrix.
Another test comes from the values of this form factor at the Callan-Treiman
point \cite{bops06} and at its soft-kaon analog \cite{bops09}. Indeed, at these particular points
the scalar form factor has a well-known value as dictated by  $SU(N_f) \times
SU(N_f)$ low energy theorems, with $N_f= 2$ and $N_f=3$, respectively.  
Combining this information with experimental results from semi-leptonic decays 
one can determine the values of the scalar form factor at these two points in the SM. 
Thus a departure from these values would be
a sign for physics beyond the SM such as right-handed quark couplings to the 
$W$ \cite{bops06,bops07} 
or charged Higgs effects, see for example the discussion in \cite{Antonelli:2008jg}(and references therein) and \cite{ddmntt09}. 
However, in order to have a reliable and accurate test of the SM one should know very precisely 
the corrections to the Callan-Treiman theorem and its soft kaon analog which are 
only exact in the soft meson limit. They are usually calculated  
in ChPT \cite{gale85}. In Ref.~\cite{bops09} the one-loop result from 
Ref.~\cite{gale85} was used and an estimation of the higher order effects was 
done since at next-to-leading order some low-energy constants (LECs) 
contribute which are not very precisely known at present.
Experimentally there has recently been  interest in trying to obtain 
the value of the scalar form factor at the Callan-Treiman point. The
three collaborations NA48 \cite{NA48mu}, KLOE \cite{KLOEmu} and KTeV \cite{KTeVmu} 
have reanalysed their data so
as to extract this value using in their analysis a dispersive representation
of the form factors proposed in Refs.~\cite{bops06, bops09}. With the current experimental 
precision the data from the last two collaborations show a good/marginal
agreement with the SM while NA48 has a $4.5 \sigma$ deviation.

The scalar form factor has been studied on the lattice. Some parameterization
of its momentum-dependence plus the knowledge of the one-loop ChPT
result at zero momentum transfer is used to extract $f_+(0)$. Here we will 
fit the lattice data from Ref.~\cite{boyle07} 
for the scalar form factor using 
a ChPT calculation at two loop order \cite{bij03}. Furthermore, we will
not only consider the scalar form factor but at the same time we will fit 
the ratio of the kaon to the pion decay constants $F_K /F_\pi$ 
\cite{amoros00,Allton08} since, as we will see, 
similar
LECs enter the two quantities. This will allow us to determine some LECs at
two-loop order (${\cal{O}}(p^6)$) 
and thus not only obtain $f_+(0)$ and 
determine $|V_{us}|$ but also  the value of the scalar form factor at 
the Callan-Treiman point and at its soft-kaon analog. Of course
one should keep in mind that we are dealing here with $SU(3)$ quantities which 
involve 
the strange quark mass. The question is whether one should consider 
the strange quark as light compared to 
the QCD scale $\Lambda \sim 200$ MeV or should it be treated as heavy. 
Related to 
that is the question whether standard $SU(3)$ ChPT which assumes that the quark 
condensate is large, is a well converging series, the 
relevant expansion parameter being in that case $(m_K/\Lambda_\chi)^2 \sim
0.4^2$. 
Also  $\bar s s$ sea quark pairs may play a significant role in chiral 
dynamics leading to different patterns of chiral symmetry breaking in $N_f=2$ 
and $N_f =3$ chiral limits \cite{mou00,dsg00}. For example, lattice QCD
seems to indicate a problem in
the  extrapolation of $F_K/F_\pi$ to its physical value when using 
$SU(3)$ ChPT to one loop order \cite{Allton08} while a fit within ``Kaon ChPT'' \cite{roe99}
where the kaon is treated as  a heavy particle leads to good agreement. 
The mass dependence of the scalar form factor has been studied within this 
scheme in Ref.~\cite{fs08}.
We will
use here standard ChPT to two loops and   
we will study the convergence of the chiral expansion.
We will also discuss the leading order ${\cal{O}}(p^4)$ LEC $L_4^r$ which
is related to the  Okubo-Zweig-Iizuka (OZI) rule violation.  

In section~\ref{chpt}, we discuss briefly the scalar form factor at two loops
in ChPT. We present the lattice calculations in section~\ref{latt} and 
discuss our fits and results in section~\ref{fit}. We conclude in section~\ref{conc}.
 
\section{ChPT to two loops} 
\label{chpt}

The strangeness changing form factors are defined from the $K \to \pi$ 
matrix element of the vector current $V_\mu =  \bar{s}\gamma_{\mu}u$
\begin{equation}
\langle \pi(p_\pi) | \bar{s}\gamma_{\mu}u | K(p_K)\rangle = 
(p_\pi+p_K)_\mu\  f_+ (t) + (p_K-p_\pi)_\mu\  f_- (t),         
\label{hadronic element}
\end{equation}
where $t\equiv q^2 =(p_K-p_\pi)^2$.
The vector form factor $f_+ (t)$ represents
the P-wave projection of the crossed channel matrix element
$\langle 0 |\bar{s}\gamma_{\mu}u | K\pi \rangle$ whereas                
the S-wave projection is described by the scalar form factor defined as
\begin{equation}
f_0 (t) = f_+ (t) + \frac{t}{m^2_{K } - m^2_{\pi }} f_-(t)~.
\label{defffactor}
\end{equation}
At zero momentum one has
\begin{equation}
f_0(0) = f_+(0)~.
 \end{equation}
These form factors 
were calculated to two loops in ChPT in Ref.~\cite{bij03}.
These authors introduced 
the quantity
\begin{equation}
\label{deftildef0}
\tilde f_0(t) = f_+(t)+\frac{t}{m_K^2-m_\pi^2}
\Big(f_-(t)+1-\frac{F_K}{F_\pi}\Big)
= f_0(t)+\frac{t}{m_K^2-m_\pi^2}\Big(1-\frac{F_K}{F_\pi}\Big)
\,.
\end{equation}
The two-loop expressions of the two decay constants $F_K$ and $F_\pi$
can be found
in Ref.~\cite{amoros00}. They involve  two $L_i$, $L_4^r$ and $L_5^r$, at ${\cal O}(p^4)$ 
and the four ${\cal O}(p^6)$ $C_i$, 
$C_{14}^r$, $C_{15}^r$, $C_{16}^r$ and $C_{17}^r$. 
Assuming that the LEC 
$L_4^r$ 
is small, which is in principle the case in the standard scenario of ChPT, 
one can expand, as usually done, the denominator in the ratio of the two decay constants
so that its contribution to order $p^4$ 
cancels and one 
is left with the contribution from $L_5^r$  
and two combinations
of three $C_i$ as detailed below:  
\begin{eqnarray}
\label{eq:Fkpi}
F_K/F_\pi &=& 1 +\frac{4}{F_\pi^2}(m_K^2-m_\pi^2) L_5^r +  \frac{8}{F_0^2} \Big[ -m_\pi^4 (C_{15}^r+2C_{17}^r)  \\
&+ &  2 m_\pi^2 m_K^2 \left(-(C_{14}^r+  C_{15}^r)+\frac{1}{2} (C_{15}^r+2C_{17}^r)\right) + 
2  m_K^4(C_{14}+C_{15}) \Big] + \delta \nonumber~,
\end{eqnarray} 
where $F_0$ is the pion decay constant in the 
chiral limit. We will come back to the discussion of this equation in Section~\ref{fit}. 
$\delta$ contains the loops and the contributions of the $L_i$
at  ${\cal O}(p^6)$.
Interestingly the dependence on these LECs is exactly the same in $f_0(t)$, see Ref.~\cite{bij03}. 
Thus the main advantage in considering $\tilde f_0$ is that this quantity
has no dependence on the $L_i^r$ at order $p^4$, only via
order $p^6$ contributions and furthermore, it only depends on the two 
${\cal O}(p^6)$ LECs $C_{12}^r$ and $C_{34}^r$. Its explicit dependence
on those is given by
\begin{eqnarray}
\label{resultfp0}
\tilde f_0(t) &=& 1-\frac{8}{F_0^2}\left(C_{12}^r+C_{34}^r\right)
\left(m_K^2-m_\pi^2\right)^2
+8\frac{t}{F_0^2}\left(2C_{12}^r+C_{34}^r\right)\left(m_K^2+m_\pi^2\right)
\nonumber\\&&
-\frac{8}{F_0^2} t^2 C_{12}^r
+\overline\Delta(t)+\Delta(0)\,,
\end{eqnarray}
where we used the notations of Ref.~\cite{bij03}.
As before, the quantities ${\overline{\Delta}}(t)$ and
$\Delta(0)$ have contributions from loops
and from the 
LECs $L_i^r$ at ${\cal{O}}(p^6)$ and
can in principle be calculated
to order $p^6$ accuracy with the knowledge of the $L_i^r$ to order $p^4$.
Parameterizations of these quantities in the physical region of $K_{\ell 3}$ 
decays can be found in Ref.~\cite{bij03}. 

Eq.~(\ref{deftildef0}) is in fact inspired by the Callan-Treiman theorem 
\cite{Callan66} which
predicts the value of $f_0(t)$ at the so called Callan-Treiman point, 
$t\equiv \Delta_{K\pi} = m_K^2-m_\pi^2$  in the 
$SU(2)\times SU(2)$ chiral limit. One has
\begin{equation}
 f_0(\Delta_{K\pi})=\frac{F_{K}}{F_{\pi}} +  
\Delta_{CT}~,
\label{eq:C} 
\end{equation}
where $\Delta_{CT}$ is a correction of $ \calo \left( m_{u,d}\right)$. It has been estimated  
within ChPT at next-to-leading order (NLO) in the isospin 
limit~\cite{gale85} with the result
\begin{equation} 
\Delta^{NLO}_{CT}=(-3.5 \pm 8.0)\cdot 10^{-3}~,
\label{Delta_CT}
\end{equation}
where the error is a conservative estimate assuming some typical corrections
of ${\cal{O}}(m_{u,d})$ and  ${\cal{O}}(m_s)$.
From Eq.~(\ref{resultfp0}) one can calculate the contribution from the ${\cal{O}}(p^6)$ LECs to 
$\Delta_{CT}$.
It reads
\begin{equation}
\Delta_{CT}|_{C_i}=\frac{16}{F_0^2} (2 C_{12}^r +C_{34}^r) m_\pi^2 (m_K^2-m_\pi^2) \, .
\label{eq:deltactci}
\end{equation}
${\cal{O}}(p^6)$ calculations \cite{bij07} using some estimates for 
the LECs $C_{12}^r$ and $C_{34}^r$ give results consistent with Eq.~(\ref{Delta_CT}). 
Strong isospin breaking as well as electromagnetic effects 
have also been evaluated \cite{bij07,kastner08}.

Another interesting quantity is the soft-kaon analog of the Callan-Treiman theorem~\cite{oehme66}
\begin{equation}
f_0(\tilde \Delta_{K \pi})= \frac{F_{\pi}}{F_{K}} + \tilde \Delta_{CT} \, ,
\label{eq:Cbis}
\end{equation}
with $\tilde \Delta_{K \pi} \equiv - \Delta_{K \pi}$.
A one loop calculation of the $SU(3)$ correction $\tilde \Delta_{CT}$
in the isospin limit \cite{gale85} gives $\tilde \Delta_{CT}=0.03$. This
is larger than its soft-pion analog $\Delta_{CT}$, see
Eq.~(\ref{Delta_CT}), by a factor $m_K^2/m_\pi^2$, however, rather small
for a first order $SU(3) \times SU(3)$ breaking effect, which is
expected to be of the order of about $25\%$. 

The value of $V_{ud}$, the first element of the CKM matrix is very 
accurately known from superallowed 
$0^+ \to 0^+$ nuclear $\beta$-decays \cite{th08}
\begin{equation}
|V_{ud}| = 0.97425 \pm 0.00022~.
\end{equation}
Combining this value with 
the experimental value of the branching ratio 
$\Gamma_{K_{l2(\gamma)}} / \Gamma_{\pi_{l2(\gamma)}}$
\cite{CKM08} and assuming the standard couplings of 
quarks to the W-boson allows to determine the ratio of the decay constants $F_K/F_\pi$.
Using 
instead the inclusive decay rate $\Gamma_{K_{Le3(\gamma)}}$ \cite{CKM08},
one obtains the value of the vector form factor at zero momentum transfer $f_+(0)$.
From these information and Eqs.~(\ref{eq:C}), (\ref{eq:Cbis}), 
one can deduce the value of the normalized form 
factor at the Callan Treiman point $C \equiv f_0(\Delta_{K \pi})/f_+(0)$ 
and at $\tilde \Delta_{K \pi}$. For the explicit formulae and more details see 
for example Ref. \cite{CKM08}. 
One has the following updated values in the SM 
\begin{eqnarray}
\label{eq:SM}
&& f_+(0)|_{SM}=0.959 \pm 0.005~, \\ \nonumber
&&   F_K /F_ \pi|_{SM}=1.192 \pm 0.006~, \\ \nonumber 
&&\ln C|_{SM}= 0.2169 \pm 0.0034 +\Delta_{CT}/f_+(0)~, \\ \nonumber
&& f_0(\tilde \Delta_{K \pi}) / f_+(0)|_{SM}= 0.8302 \pm 0.0074 +\tilde{\Delta}_{CT}/f_+(0)~.
\end{eqnarray}
Deviations from these SM predictions would thus be a sign of new physics. 
For example at NLO within the minimal not-quite decoupling
electroweak low-energy effective theory (LEET) \cite{hs04}, 
in the light quark sector one has
two combinations of parameters of spurionic origin describing the
couplings of quarks to the $W$-boson to be determined from
experiment~\cite{stern06,bops09}. While the knowledge of the scalar
form factor at the CT point measures one combination, its knowledge at
$\tilde \Delta_{K \pi}$ measures the other one. A precise determination of 
$ \Delta_{CT}$ and
$\tilde \Delta_{CT}$ would thus help to settle the issue of the
presence of right-handed couplings of quarks to the $W$-boson.

\begin{table}[h!]
\begin{center}
\begin{tabular}{|l||c|c|c|c|}
\hline
&Fit 10 \cite{fit10}&$\pi K$ Roy Steiner~\cite{buett03}& Prelim. Fit All(*) \cite{bijjem10}
& Lattice \cite{Allton08}\\
& set a   & &set b&\\
\hline
$10^3 L_1^r$&0.432 & $1.05 \pm 0.12$ &$0.99\pm0.13$ &$-$ \\
$10^3L_2^r$& 0.735 &$1.32 \pm 0.03$&$0.60 \pm 0.21$ &$-$\\
$10^3 L_3^r$&$-2.35$ &$-4.53 \pm 0.14$ &$-3.08\pm0.47$ &$-$\\
$10^3 L_4^r$& 0 &$0.53 \pm 0.39$ &$0.70\pm0.66$ &$0.33(0.13)$\\
$10^3 L_5^r$& 0.97 &$3.19 \pm 2.40$ &$0.56\pm0.11$ &$0.93(0.073)$\\
$10^3 L_6^r$&0 &  &$0.14\pm0.70$ &- \\
$10^3L_7^r$&$-0.31$ &  &$ -0.21\pm0.15$&-\\
$10^3 L_8^r$& 0.6& &$0.38\pm0.17$ &- \\
$10^3 (2 L_6^r-L_4^r)$ &&&  &0.032 (0.062) \\
$10^3 (2 L_8^r-L_5^r)$ &&& &0.050(0.043) \\
\hline
\end{tabular}
\end{center}
\caption {\label{table:lec} ${\cal{O}}(p^4)$ LECs at a scale $\mu=0.77$~GeV.}
\end{table}

In order to have a very precise
determination of $f_+(0)$ as well as $\Delta_{CT}$ and $\tilde \Delta_{CT}$, 
one needs to have a very precise determination
of all the LECs $L_i^r$ and $C_i^r$ which enter 
Eqs.~(\ref{eq:Fkpi}, \ref{resultfp0}). 
\begin{itemize}
\item The $L_i$ have been determined in Ref.~\cite{fit10} 
from a fit to the masses and to $K_{l4}$-decay data from the E865 experiment,  
assuming that $L_4^r$ and $L_6^r$ are $1/N_c$ suppressed and using
$F_K/F_\pi=1.22$~{\footnote{In this fit some of the $C_i$ are taken from resonance 
saturation, the others are set to zero, see Ref.~\cite{amoros00}}}. 
Matching the dispersive results for the subthreshold expansion parameters
of $\pi K$ scattering with their chiral expansion at order $p^4$ \cite{buett03} 
leads to somewhat different results, especially $L_4^r$ is suggestive 
of a significant violation of the OZI rule in the scalar sector, see
Table~\ref{table:lec}. 
This is in agreement with a determination of some of the LECs
in an analysis of $J/\psi$ decays into vector mesons and two 
pseudoscalars \cite{lae06}.

\item In Ref.~\cite{ecker89} it was shown that it was possible to reproduce the values of
the $L_i$ in terms of properties of the light meson resonances (masses and
coupling constants). 
The idea of using resonance saturation also for the ${\cal{O}}(p^6)$ LECs was thus taken up 
and the $C_i$ are presently mostly estimated in that framework \cite{Cirigliano06,ceekpp05,Ecker07}.
There are, however, a few problems. First the scale at which they are obtained is not known. It is 
usually assumed  to be given by the lightest scalar nonet that 
survives in the large $N_c$ limit, $M_S=1.48$~GeV. The value at
another scale, typically the $\rho$ mass scale, is 
obtained using  renormalization group equations. Furthermore, a test of the 
naturalness of the $C_i$ \cite{karol06} shows that some of them are in fact not
dominated by the resonance contributions. Also the LECs we are
interested in have important contributions from the scalar sector 
where one knows that the OZI rule is strongly violated and where the
presence of the wide scalar $\sigma$ and $\kappa$ mesons makes the calculation
in terms of tree level diagrams from a resonance Lagrangian not really 
appropriate. Considering more specifically $C_{12}^r$ and $C_{34}^r$,
several calculations have been performed
based on the study of the scalar form factor with $\Delta S=0$ \cite{bijdh03} or 
$\Delta S=1$ \cite{jam04,ber07}. 
In the literature these two LECs, Eq.~(\ref{resultfp0}), lie in the range
$-10^{-3}$ GeV$^{-2}$ to a few $ 10^{-4}$ GeV$^{-2}$. The four other 
$\calo (p^6)$ LECs ($C_{14}^r$, $C_{15}^r$  $C_{16}^r$ and $C_{17}^r$), Eq.~(\ref{eq:Fkpi}), 
needed in our study are not very well known. In Ref.~\cite{jiang09} where  the $C_i$ have  been 
recently determined within a quark model, one finds
$C_{15}^r=C_{16}^r=0$, $ C_{17}^r=0.01 \cdot 10^{-3}$ GeV$^{-2}$ 
and $C_{14}^r=-0.83 \cdot 10^{-3}$ GeV$^{-2}$ which is smaller than
what is found in resonance saturation  $C_{14}^r=-4.3 \cdot 10^{-3}$ GeV$^{-2}$. 
\end{itemize}
With the progress of lattice QCD it becomes also possible to
extract the LECs from a chiral extrapolation of the lattice data.
Already some of the ${\cal{O}}(p^4)$ ones have been obtained mostly
within $SU(2)$ ($l_i$). Relations between the $SU(2)$ and the $SU(3)$ LECs 
allows to 
determine the $L_i$ from the $l_i$ \cite{gale85a,ghis07} (for similar relations between the
$C_i$ see Ref.\cite{Christoph09}). 
Results from the RBC/UKQCD collaboration are shown in Table~\ref{table:lec}.
As can be seen from this table
most of the ${\cal{O}}(p^4)$ LECs are still not well enough determined for a very precise
test of the SM. 
A global fit of all the low-energy constants of Chiral Perturbation theory
at next-to-next-to-leading order currently performed \cite{Ilaria09,bijjem10} will hopefully 
help to settle the values of these LECs much more
precisely. Some preliminary results \cite{bijjem10} which differ from fit~10 by using
some more recent data, by letting $L_4^r$ and $L_6^r$ free and  by adding some constraints 
from $\pi K$ scattering show better agreement with the analysis of Ref.~\cite{buett03} 
as the comparison between the second and third column of Table~\ref{table:lec} shows.

\section{Lattice} 
\label{latt} 
Following the pioneering work of Ref. \cite{bice05} 
different collaborations have extracted the vector form factor at zero momentum transfer either with 
$N_f=2$ \cite{JLQCD05,RBC06,QCDSF07,ETMC09} 
or $N_f=2+1$ \cite{boyle07} flavours. The idea is to first evaluate the scalar form factor
$f_0(t)$ at the momentum transfer $t_{{\rm  {max}}} =(m_K-m_\pi)^2$. This can be very efficiently done
calculating a double ratio of three-point correlation functions \cite{bice05}.
Then a phenomenologically motivated interpolation is performed up to zero momentum transfer {\footnote{A new technique has been developed in Ref. \cite{bo07} which will allow to directly simulate at $t=0$ on the lattice.}} and the 
Ademollo-Gatto theorem is used to obtain a rather precise value for $f_+(0)$.
Let us consider the chiral expansion of $f_+(0)$
\begin{equation}
f_+(0)= 1 +f_2 + f_4 +\cdots~, 
\end{equation}
where $f_n = {\cal O}((m_{K,\pi}/(4 \pi F_\pi))^n)$ and the first term is equal 
to one due to gauge invariance. 
The Ademollo-Gatto theorem \cite{Ademollo64} states that the deviation from unity
of $f_+(0)$ is predicted to be second order in $SU(3)$ symmetry breaking, 
{\it i.e.} of order $(m_s-\hat m)^2$, where $m_s$ and $\hat m$ are the strange 
and the average of the $u,d$ quark masses, respectively 
{\footnote {Note, however, that despite this theorem the 
light quark mass difference $m_u \neq m_d$ can modify $f_+(0)$ to first order.}} so  that
the ${\cal {O}}(p^2)$ term $f_2$ in the chiral expansion of $f_+(0)$ is free of any LECs.
The different collaborations generally take this term from a one-loop ChPT 
calculation \cite{gale85}
\begin{equation}
f_2=-0.0227~,
\end{equation}
obtained for pion, kaon and eta masses taken at their physical values and in
the isospin limit
and determine the difference 
\begin{equation}
\Delta f= f_+(0)-1-f_2~.
\label{eq:f2}
\end{equation}
This difference contains of course all terms starting at the order ${\cal {O}}(p^6)$. Also used is the partially quenched expression derived in Ref. \cite{be05}. An expression for $f_2$  using NLO SU(2) ChPT can be found in Ref. \cite{fs08}. 
The first determination of $\Delta f $ in a quark model framework gave $\Delta f=-0.016(8)$
\cite{LR84}. 
   
The RBC/UKQCD collaboration for example  
\cite{boyle07,Allton08} simulates with $N_f=2+1$ flavors of dynamical domain 
wall quarks. In order to determine
$f_+(0)$, they performed a simultaneous fit to both the $t$ 
and quark mass dependences using the ansatz
\begin{equation}
f_0(t,m_\pi^2,m_K^2) =\frac{1 +f_2 +(m_K^2-m_\pi^2)^2(A_0+A_1(m_K^2+m_\pi^2))}{1-t/(M_0+M_1(m_K^2+m_\pi^2))^2}~.
\end{equation}
This formula motivated by the Ademollo-Gatto theorem 
has four fit parameters $A_0, \,A_1$, $M_0, \,M_1$ and $f_2$
is the NLO term, Eq.~(\ref{eq:f2}). They have also used a second order 
Taylor expansion as parameterization of the $t$-dependence of the form factor
{\footnote{This 
parameterization and the pole one are usually assumed either in 
lattice calculations or in most of the experimental analyses. 
One should note, however, that the pole parameterization
has no real physical motivation in the case of the scalar form factor. Also it has been shown \cite{KTeVmu,bops09} that in
order to get a very precise parameterization of the scalar form factor  in the physical
region of $K_{l3}$ decay ($m_\ell^2 < t < t_{max}$), an
expansion up to third order had to be done.}}.
They obtain
\begin{equation}
f_+(0)=0.9644 \pm 0.0033_{\rm stat} \pm 0.0037_{\rm syst}~.
\end{equation}
The same collaboration has also extracted the ratio $F_K / F_\pi$ \cite{Allton08}
\begin{equation}
F_K /F_ \pi = 1.205 \pm 0.018_{\rm stat} \pm 0.062_{\rm syst}~.
\end{equation}
A summary of other lattice results can be found in Refs.~\cite{CKM08, FLAG}. In the 
unquenched simulations they fall in the range between $1.189$ and $1.218$ for the
central value of $F_K/F_\pi$ and between 0.956 and 0.968 for the one
of $f_+(0)$. While the errors on the former are very small, they are
larger on the latter. 
All these numbers should be compared to the Standard Model values, Eq.~(\ref{eq:SM}).

\section{Chiral Extrapolation} 
\label{fit} 
We now turn to the central point of the paper, namely the 
chiral extrapolation of the lattice data on $F_K/F_\pi$ and $f_+(0)$ based on 
the  two-loop ChPT calculations \cite{bij03}. We use the 
results from the RBC/UKQCD collaboration since this is the only collaboration which has calculated both
these quantities with $N_f=2+1$ flavors. We take the data performed 
on the $24^3 \times 64$ volume with
an inverse lattice spacing of $a^{-1} =1.73(3)$ GeV and a simulated strange
quark mass, $a m_s =0.04$ close to its physical value. 
We do not correct
for finite volume effects (FV) or lattice artefacts (LA). They have been 
estimated
for $F_K/F_\pi$ \cite{Allton08} where the error bars they quote for these effects are roughly equal
(FV) or even larger (LA) than the statistical ones. We only included the statistical errors in our fits. Also we did not include the
correlations between $F_K$ and $F_\pi$ since they are not available.
Lattice results have been obtained for four values 
of the light quark masses which correspond to pion (first number in parenthesis) and kaon masses (second 
number) equal to ($0.329, 0.575$)~GeV (set (I)) 
($0.416, 0.604$)~GeV (set (II)), ($0.556,0.663$)~GeV (set (III)) and ($0.671,0.719$)~GeV. 
Clearly, ChPT cannot be valid at too high 
pion and kaon masses so we completely discard the last set in our fits and mostly use sets~(I) and (II). 
For each pion mass they 
have calculated the scalar form factor at five values of $t$ going from 
$\sim -0.4$ GeV$^2$ to $t_{\rm max}$. Again for
the fits we only use the three smallest absolute values of $t$.

\begin{table}[h!]
\begin{center}
\begin{tabular}{|l||c|c|c|c|c|c|}
\hline
&Fit I & Fit II &Fit III & Fit IV& Fit V& Fit VI\\
\hline
$ C_{12}$&$5.77 \pm 0.56$ &$7.84 \pm0.58 $&$ 4.69 \pm 0.95$ & $5.74 \pm 0.95  $&$ 4.69 \pm 0.56$ &$4.43 \pm 0.88$\\
$ C_{34}$&$2.54 \pm 0.43 $ &$ -1.28\pm 0.44$&$ 3.76  \pm 0.95$ &$ 1.07 \pm 0.96 $ &$3.76 \pm 0.43$ &$3.50 \pm 0.94$\\
$C_{14}$&$0^*$ &$0^*$&$0.65 \pm 1.38$ &$0.71\pm 1.42  $&$0.65^*$&$-0.93 \pm 0.67$ \\
$ 2 C_{17}$&$0^*$ &$0^*$&$ 0.31 \pm 3.31$&$ 1.92 \pm 3.36 $ &$0.31^*$&$4.16 \pm 1.56$\\
$F_0$  &$89.8 \pm 0.1 $ &$ 69.2\pm 0.0 $&$ 89.8 \pm 0.1 $ & $ 69.3 \pm 0.0 $ &$89.8^* $&$ 89.8\pm 0.1 $ \\
$ f_+(0)$ &0.956 & 0.963& 0.956 &0.961 & 0.956 & 0.958\\
$F_K/F_\pi$& 1.20 & 1.19&1.20 & 1.19 &1.20& 1.19\\
$ \ln C$ &0.22 &$ 0.20 $&0.22 &0.21 &0.22&0.21\\
$ f_0(\tilde \Delta_{K \pi})$ &0.75&0.75  &0.75&0.76&0.75&0.77 \\
$10^3\Delta_{CT}$& 1.00 &$ -2.14$&0.27 &$-3.65$ &0.18& $-0.32$\\
$10^2\tilde \Delta_{CT}$&$-9.00$ & $-9.86$&$-8.24$ & $-8.18$&$-8.11$&$-7.03$\\
$10^3\lambda_0$ &18.08 &17.77&18.24 & 17.66&18.18&16.71\\
$\chi^2$ &1.40/4 &0.96/4 &1.67/4 &1.29/4&3.01/4&4.8/7\\
\hline
\end{tabular}
\end{center}
\caption {\label{table:fit}  Result of the fits to the lattice data. 
The first five quantities are the parameters of the fits. The star denotes
an input quantity. The $C_i$'s are in units
$10^{-4}$ GeV$^{-2}$ and  $F_0$ is in MeV. In Fits (I) and (II) 
the lattice data on $F_K/F_\pi$ are not included.} 
\end{table}

A quantity $\calo$ at two loops  has typically the following form after renormalization
of the pion decay constant
\begin{equation}
\calo (m_\pi,m_K,m_\eta)= \calo_{LO} +\frac{\calo_{\rm
    NLO}}{F_\pi^2}+\frac{\calo_{\rm NNLO}}{F_0^4}~,
\label{eq:Op}
\end{equation}
where $\calo_{LO}$, $\calo_{\rm NLO}$ and $\calo_{\rm NNLO}$ are 
the contribution at leading order (LO), next-to-leading order (NLO) and
next-to-next-to-leading order (NNLO), respectively. Here,
$F_\pi$ is the pion decay constant calculated at $\calo (p^4)$ at
the value of the pion mass and of the kaon mass under consideration
and $F_0$ is the pion decay constant in the chiral $SU(3)$ limit. 
When working at the physical pion and kaon masses one usually replaces everywhere
the decay constant by its physical value,
since the difference is of higher order. This is the procedure which has been
used to determine the LECs, set~(a) and (b) of Table~\ref{table:lec}. If this
is mostly justified for set~(a) where the difference between $F_\pi$ and $F_0$
is small, this is more questionable for set~(b) where $F_0=67.1$~MeV but allows, of
course, for a better convergence of the chiral series. Also 
going away from the physical point the difference between $F_\pi$ and $F_0$ might 
become again too large for this procedure to be entirely satisfying. Here we will
just replace $F_0$ by the physical value of $F_\pi$ in the NNLO term in order to be 
consistent with the determination of the LECs. Also to be consistent
with their determination we will use Eq.~(\ref{eq:Fkpi}) for determining $F_K/F_\pi$.
Again if this
is justified for set (a) where the convergence of this quantity is rather good
as we will see below, this is more questionable for set (b). 
In the expression Eq.~(\ref{eq:Op}),  the mass of the $\eta$ enters the NLO and NNLO terms. In 
the calculation of $\calo_{\rm NLO}$  its NLO expression 
is used while in $\calo_{\rm NNLO}$ the $\eta$ mass is given by the Gell-Mann-Okubo relation.  

We have performed several fits to the lattice data and determined from
these fits results for $f_+(0)$, the slope of the scalar form-factor at
zero momentum transfer $\lambda_0$, $F_K/F_\pi$, $\Delta_{CT}$ and 
$\tilde \Delta_{CT}$. We have taken
$F_0$ as a parameter of the fit using the value of the physical pion decay constant
as input.  
Apart from Fit~(VI) they are done with  the two lattice data sets with 
the smallest pion values, sets~(I) and (II).  All the fits are done for the 
three smallest absolute values  of $t$. The values of the LECs $L_i$ are taken 
from  sets ~(a) and (b) of Table~\ref{table:lec}. These sets correspond to 
a value of $m_s/\hat m=24$. In Ref.~\cite{Ilaria09}
another preliminary set is given corresponding to a somewhat
larger value  $m_s/\hat m=27.8$ as obtained by MILC and HPQCD/UKQCD. 
It leads to an even smaller value of $F_0=62.7$ MeV and will not be discussed here.
The results of the fits are given in Table~\ref{table:fit}. The one from
this other set are comparable to the one of set (b). 
\begin{itemize}
\item Fits~(I) and (II) 
are  three parameter fits of $f_0(t)$
using sets~(a) and (b) respectively. The $C_i$ are the
one used in the determination of the $\calo (p^4)$ LECs, fit~10, namely 
$C_{14}^r=C_{15}^r=C_{16}^r=C_{17}^r=0$. 
For  set (a) $F_K/F_\pi=1.22$ whereas for  set (b) $F_K/F_\pi=1.19$.
Slightly different values are given in the table for set (a) since, as explained below 
Eq.~(\ref{eq:Op}) we did not use the 
physical value of $F_\pi$ in the calculation of this quantity in the NLO term but
rather its NLO expression.  

\item Fits~(III) and (IV) are combined fits of $F_K/F_\pi$ and $f_0(t)$ 
using sets~(a) and (b), respectively, as in the previous fits  but now the combinations
$C_{14}^r+C_{15}^r$ and $C_{15}^r+ 2C_{17}^r $ which appear in $F_K/F_\pi$ are left free. 
Since we need to determine $F_0$, we, in principle, need to know $C_{16}^r$ and the combination
$C_{15}^r-2C_{16}^r$. We will assume them equal to zero, this is consistent with
the results in Ref.~\cite{jiang09}. Thus we do in fact determine $C_{14}^r$ and $C_{17}^r$.

\item Fit~(V): here we fix the combinations $C_{14}^r+C_{15}^r$ and
  $C_{15}^r+2 C_{17}^r$ from Fit~(III) and 
we fit the quantity $\tilde f_0(t)$. 

\item Fit~(VI) is the same as Fit~(III) but 
 with the lattice data for $f_0(t)$ from set~(III) 
also included.

\end{itemize}

\begin{figure}[h!]
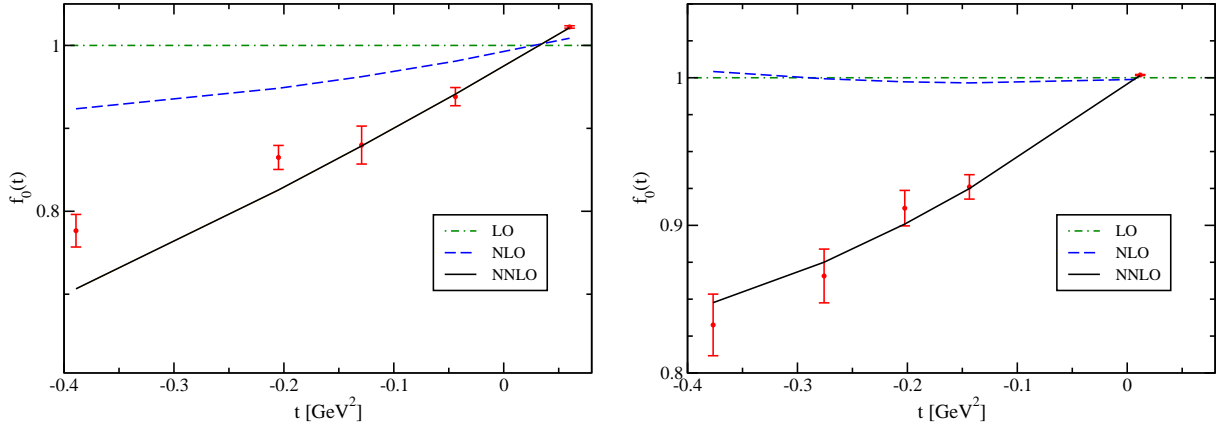

\begin{center}
\includegraphics*[scale=0.325]{fitvii1.eps}
\hspace{0.4cm}\includegraphics*[scale=0.325]{fitvii3.eps}
\caption{Momentum-dependence of the scalar form factor. The results of
Fit~(VI) are displayed (solid line) for set~(I) (left panel) and
set~(III) (right panel). The convergence of the chiral
expansion is also displayed: the dash-dotted line is the result at LO, the
dashed line displays the one up to NLO.}
\label{results}
\end{center}
\end{figure}

As can be seen from Table~\ref{table:fit}, we obtain very good fits of the lattice data.  
Fits~(I) and (II), however, do not reproduce well the two lattice points for
$F_K/F_\pi$ from sets~(I) and (II).
Fits~(III) and (IV) which correspond to two very different values of $L_4^r$ are comparably good, 
but an order of magnitude larger value of  
 $ C_{17}^r$
is in fact needed in order to
compensate for the larger value of $L_4^r$ in Fit~(IV)  compared to Fit~(III). 
$C_{14}^r$ and $ C_{17}^r$ are at least an order of
magnitude 
smaller than what is expected from  resonance saturation in the scalar sector which 
leads to typical values $\sim 10^{-3}$. One has for
example \cite{amoros00}
\begin{equation}
C_{14} \sim \frac {c_d c_m d_m}{M_S^4} \sim -4.3 \cdot 10^{-3} \, {\rm {GeV}}^{-2}
\end{equation}
where $ c_d$, $c_m$ and $d_m$ are coefficients of the scalar chiral Lagrangian.
$M_S$ and $d_m$ are obtained from the masses of the scalars $K^*_0(1430)$
and $a_0 (980)$ and $c_m=0.042$~GeV and $c_d =0.032$~GeV. 
The results of Fit~(VI)  do not differ much from 
Fit~(III), only  $C_{14}$ and $ C_{17}$ are
larger in absolute value and the slope of the scalar form factor is somewhat smaller.
 This fit is shown on Fig.~\ref{results} 
for sets~(I) and (III).  Even though we only fit the 
three smallest points in absolute
value, the $t$-dependence of set~(III) is remarkably well reproduced by
ChPT to two loops.      

Fitting $f_0(t)$ leads to  strong anticorrelations between $C_{12}$ and $C_{34}$
on the one hand and $C_{14}+C_{15}$ and $C_{15}+2 C_{17}$ on the other one, typically of the
order of $-0.8$ while in Fit~(V) the correlations between $C_{12}$ and $C_{34}$ are reduced by 
a factor of two. Also a
comparison of Fits~(III) and (V) shows that the error bars on these two LECs are smaller 
in the latter case. 
Thus a rather good determination of the LECs $C_{12}$ and $C_{34}$ is obtained by fitting the 
function $\tilde f_0$. Their order of magnitude is the one expected
from resonance saturation. Note that the value obtained for $C_{12}+
C_{34}$ is rather independent of the fits within one set, one gets $\sim 8 \cdot
10^{-4}$ GeV$^{-2}$ for set~(a) and  $\sim 6 \cdot
10^{-4}$ GeV$^{-2}$ for set~(b).

The results for  $F_K/F_\pi$, $f_+(0)$ and $\ln C$  are consistent
with the values obtained assuming the standard quark couplings to the $W$-boson, 
Eq.~(\ref{eq:SM}). 
We refrain to give error bars here since one should have a 
more precise knowledge of the $L_i$ as well as lattice data at lower pion and
kaon masses to really be able to pin down these quantities very precisely.
Difference between the various sets  gives an idea of the errors.    
The value of $\lambda_0$  turns out to be rather large compared 
to the experimental results, the lattice determination of  Ref.~\cite{ETMC09}
or to what is obtained from the formula obtained in a dispersive parameterization
of the form factor \cite{bops06,bops09}
\begin{equation}
\lambda_0= \frac{m_\pi^2}{(m_K^2-m_\pi^2)}(\ln C- G(0)) , \,\, \,\, \,\, \,\,
G(0)=0.0398 \pm 0.0044 \, ,
\label{la0}
\end{equation}
where $G(t)$ is a dispersive integral of the phase of the form factor which is
identified in the elastic region with the s-wave, 
$I=1/2$ $K \pi$ scattering phase according to Watson's theorem. 
In the analysis \cite{bops06,bops09} it was taken
from \cite{buett03} where a matching of the solution of  the Roy-Steiner equations
with the $K \pi \to K \pi $ , $\pi \pi \to K \bar K$ and $\pi \pi \to\pi \pi$
scattering data available at higher energies has been performed. 
Note that in this analysis the LECs obtained, second column of Table \ref{table:lec},  are more consistent
with the values used in Fit~(IV), especially a large violation of 
the OZI rule was found. This large value of $\lambda_0$ can be traced back to the too large
value of the combination $2 C_{12}+ C_{34}$ which enters its expression within 
ChPT, see Eq.~(\ref{resultfp0}). It is however compensated by
a small curvature $\lambda_0'$ leading to a value of the scalar form factor at the 
Callan-Treiman point in agreement with the SM value. Typically one obtains 
$\lambda_0' \sim 1 \cdot10^{-4}$ instead of  $\sim 6 \cdot10^{-4}$  as expected from experiments and dispersive analyses \cite{jop06,bops06}. 
Again 
$C_{12}$ has a too large positive value. Stringent constraints on slope and
curvature have recently been obtained using the method of unitarity bounds
\cite{capr09}.

Let us study the convergence of the results. In Fig.~\ref{results} is
shown $f_0(t)$ as obtained in Fit (VI) at  LO (dot dashed line),
NLO (dashed line) and NNLO (full line). On the
left-hand-side (LHS) set~(I) is displayed  and on the 
right-hand-side (RHS) set (III), 
in order to compare the dependence on the pion and the kaon masses.   
Clearly, as expected,
the convergence of $f_0(t)$ worsens as one increases the absolute value of $t$
(LHS, set~(I)) and as one
increases $m_\pi$ and $m_K$.
At the physical pion and kaon masses one has from Fit~(III), 
\begin{eqnarray}
f_+(0)&=&1-0.019 -0.026 + \ldots , \nonumber \\
 F_K/F_\pi&=&1+0.140 +0.061  + \ldots , \nonumber \\
 f_0(\Delta_{K \pi}) &=&1+0.139+0.063  + \ldots  , \\
\Delta_{CT}&=&0-0.0025+0.0028 + \ldots , \nonumber \\
\tilde \Delta_{CT}&=&0+0.024 -0.106 + \ldots , \nonumber
\end{eqnarray}
and from Fit~(IV)
\begin{eqnarray}
f_+(0)&=&1-0.019 -0.019 + \ldots , \nonumber \\
 F_K/F_\pi&=&1+0.113 +0.081  + \ldots , \nonumber \\
 f_0(\Delta_{K \pi}) &=&1+0.110+0.081  + \ldots  , \nonumber \\
\Delta_{CT}&=&0-0.0033 -0.0003  + \ldots , \nonumber \\
\tilde \Delta_{CT}&=&0+0.021 -0.103 + \ldots , 
\end{eqnarray}
where the first, second and third terms are the ${\cal{O}}(p^2)$, ${\cal{O}}(p^4)$ 
and ${\cal{O}}(p^6)$ contributions, respectively, and the ellipses denote
terms of order $p^8$ and higher. Note that by definition $\Delta_{CT}$ and $\tilde \Delta_{CT}$ have no LO contribution. The convergence is rather good/not very good for $F_K/F_\pi$
and $f_0(\Delta_{K \pi})$ for set (a) and set (b) respectively while the one for  
$\Delta_{CT}$ is good for set (b) and not for set (a). One should however keep in mind that
the NLO correction for this last quantity is small being an $SU(2) \times SU(2)$ one. Also the NNLO contribution of $\Delta_{CT}$
is of the expected size of the corrections, Eq.~(\ref{Delta_CT}).
The convergence of $f_+(0)$
and  $\tilde \Delta_{CT}$ is bad whatever the set.    
However the 
convergence looks again worth than it is in reality. Indeed for both quantities 
the contribution at NLO is smaller than naively expected. For $f_+(0)$
this is essentially due to the Ademollo-Gatto theorem, as we have seen in the previous section. 
In both cases the 
NNLO term is of the expected size.
Let us look in a bit more details at the diverse contributions for 
$f_+(0)$, $F_K/F_\pi$ and $f_0(\Delta_{K \pi})$. One has for Fit (III)
\begin{eqnarray}
f_+(0)&=&1+(-0.019+0.000)+(0.012-0.003-0.034) +\ldots , \nonumber \\
 F_K/F_\pi&=&1+(0.057+0.083) + (-0.005 +0.045+0.021) +\ldots ,\nonumber \\
f_0(\Delta_{K \pi})&=&1+(0.055+0.083)+(-0.001+0.047+0.017) + \ldots ,
\end{eqnarray}
and for Fit (IV) 
\begin{eqnarray}
f_+(0)&=&1+(-0.027+0.008)+(0.012-0.002-0.029) +\ldots , \nonumber \\
 F_K/F_\pi&=&1+(0.086+0.027) + (-0.005 +0.078+0.009) +\ldots ,\nonumber \\
f_0(\Delta_{K \pi})&=&1+(0.083+0.026)+(-0.001+0.063+0.019) + \ldots . 
\end{eqnarray}
The first brackets give the contribution from the loops and the
$L_i$ at fourth order and the second brackets represent the one at sixth
order from the two-loops, the one-loop with one 
$L_i$ insertion plus tree graphs with two $L_i$ and the tree graphs $\sim C_i$, in order. 
One sees that the large contribution of $f_+(0)$
at NNLO is due to big corrections of the  dimension six operators, as was the case
for the slope and the curvature, see the discussion before. It could  be that
the corresponding LECs $C_i$ are larger than they are in nature mocking up
some higher order effects. 
The contributions from
the two-loop and the one-loop $\sim L_i$ topologies do 
converge. In the case of $F_K/F_\pi$  and $f_0(\Delta_{K \pi})$ it is the 
terms proportional to $L_i$ which are responsible
for their not so  good convergence in the case of set (b), explaining the difference between 
the two sets. 
Let us consider also the convergence of  $F_K/F_\pi$ at larger pion
and kaon masses. One has for Fit~(IV)
\begin{eqnarray}
F_K/F_\pi&=&1+0.043+0.093 + \ldots =1.136 +\ldots , \,\,\,\,\, \,\,\, \rm{set~(I)} \nonumber \\
&=&1+0.023+0.076 +\ldots =1.099 +\ldots , \,\,\,\,\, \,\,\, \rm{set~(II).}
\end{eqnarray}
For comparison the lattice data are:
\begin{eqnarray}
F_K/F_\pi&=&1.134 \pm 0.011 , \,\,\,\,\, \,\,\, \rm{set~(I)} \nonumber \\
&=&1.101 \pm 0.010, \,\,\,\,\, \,\,\, \rm{set~(II).}
\end{eqnarray}
As already stated for the scalar form factor and as expected, the convergence 
gets worse when increasing the values of $m_\pi$ and $m_K$. This bad convergence
could be an artefact of the use of lattice data obtained at still too high pion and kaon masses for ChPT to really be valid. 

\section{Conclusion} \label{conc}

We have done here a first exploratory study using a two-loop ChPT calculation
to fit the lattice data.  Certainly finite volume effects  for example should 
be taken into account in a more refined treatment. However, before this can be 
done, a better knowledge of the $L_i$ are
necessary and more lattice data at smaller  masses are needed. This is  
important for
checking the convergence of the $SU(3)$ ChPT calculations as well as for
a more precise determination of the  quantities studied here. Also
if the result of set (a) is not very sensitive to the treatment of the
NNLO term, see discussion below Eq.~(\ref{eq:Op}), this is clearly not the
case for set (b) and our results here are certainly not the final ones.
Indeed, if large values for $L_4^r$ and $L_6^r$ as expected from a large violation of the OZI rule  
were confirmed in the future then 
the use of standard ChPT as done here would not really be appropriate.  A way of solving the 
problem could be for example to work within resummed ChPT \cite{d07}. A study along
this line is in progress \cite{bdt10}.

\subsection*{Acknowledgements}
We are extremely grateful to Johan Bijnens for giving us his programs for
calculating the scalar and vector form factors.  
Without his help this work would not have been possible.  
We would like to thank the RBC/UKQCD collaboration for providing us with the lattice data on the scalar form factor. We are grateful to  Sebastien
Descotes-Genon, Christoph Haefeli, Ilaria Jemos, Andreas J\"uttner  and Ulf-G. Mei{\ss}ner
for interesting discussions and useful comments. This work has been supported in part by 
the EU contract
MRTN-CT-2006-035482 ("Flavianet"), the European
Community-Research Infrastructure Integrating Activity "Study of Strongly Interaction Matter" (acronym Hadron-Physics2, Grant Agreement n. 227431) under the Seventh Framework Programme of the
EU, by MEC (Spain) under grant FPA2007-60323 and by the
Spanish Consolider-Ingenio 2010 Programme CPAN (CSD2007-00042).

\end{document}